\documentclass[aps,prd,twocolumn,groupedaddress,nofootinbib,longbibliography]{revtex4-2}

\usepackage{graphicx}
\usepackage{amsmath,amssymb,amsfonts}
\usepackage[colorlinks,citecolor=blue,linkcolor=blue,urlcolor=blue]{hyperref}
\usepackage[applemac]{inputenc}

\newcommand{\be}{\begin{equation}}
\newcommand{\ee}{\end{equation}}
\usepackage[english]{babel}

\newcommand{\capo}[1]{\subsection*{#1}}

\usepackage[autostyle, english = british]{csquotes}

\begin{document}

\title{\LARGE The Relational Interpretation}
\author{Carlo Rovelli}
\affiliation{Aix Marseille University, Universit\'e de Toulon, CNRS, CPT, 13288 Marseille, France.\\ Perimeter Institute, 31 Caroline Street North, Waterloo, Ontario, Canada, N2L 2Y5.\\ The Rotman Institute of Philosophy, 1151 Richmond St.~N London, Ontario, Canada, N6A 5B7.}

\begin{abstract}
\centerline{To appear as a Chapter in the \emph{Oxford Handbook of Quantum Interpretations}.}
\end{abstract}

\maketitle

\noindent The relational interpretation (or RQM, for Relational Quantum Mechanics) solves the measurement problem by considering an ontology of sparse \emph{relative} events, or facts. Events are realized in interactions between \emph{any} two physical systems and are relative to these systems. RQM's technical core is the realisation that quantum transition amplitudes determine physical probabilities only when their arguments are facts relative to the same system.  The relativity of facts can be neglected in the approximation where decoherence hides interference, thus making facts approximately stable.  

\capo{Historical roots}

In his  celebrated 1926 paper \cite{Schrodinger:1926fk}, Erwin Schr\"odinger introduced the wave function $\psi$ and computed the spectrum of hydrogen from first principles.   This spectrum, however, had already been computed from first principles by Pauli four month \emph{earlier}  \cite{Pauli1926}, using the theory that emerged  from Werner Heisenberg's 1925 breakthrough \cite{Heisenberg1925a}, based on the equation
\be
qp-pq=i\hbar,
\label{1.2}
\ee
with no reference to $\psi$.
The theory we call ``quantum mechanics", in fact, had already evolved into its current {full} set of equations in the series of articles by Born, Jordan and Heisenberg himself \cite{Born1925b,Born1926}. Dirac, equally inspired by Heisenberg's breakthrough, got to the same structure independently in 1925, the year \emph{before} Schr\"odinger's work, in a paper titled ``The fundamental equations of quantum mechanics" \cite{Dirac1925}.   (See \cite{Fedak2009,Waerden1967} for a detailed historical account.)  Properly, the only Nobel Prize with the motivation ``for the creation of quantum mechanics" was assigned to Heisenberg. 

So, what did Schr\"odinger achieve in 1926?   With hindsight, he took a {technical}  and a  {conceptual} step. The {technical} step was to translate the unfamiliar algebraic language of quantum theory into a familiar one: differential equations. This brought the novel ethereal quantum theory down to the level of the average theoretical physicist.   The {conceptual} step was to introduce the notion of ``wave function", which soon evolved into the general notion of ``quantum state", $\psi$, endowing it with ontological weight.    

The Relational Interpretation of Quantum Mechanics, or RQM, is based on the idea that this {conceptual} step, \emph{which doesn't add anything to the predictive power of the theory}, was misleading: we are paying the price for the confusion it generated.   

The mistaken idea is that the quantum state $\psi$ represents the ``actual stuff" described by quantum mechanics. This idea has pervaded later thinking about the theory, fostered by the toxic habit of introducing students to quantum theory in the form of  Schr\"odinger's ``wave mechanics",  thus betraying history,  logic, and reasonableness. 

The true founders of quantum mechanics saw immediately the mistakes in this conceptual step. Heisenberg was vocal in pointing them out \cite{Kumar2008}:  First, Schr\"odinger's ground for considering $\psi$ to be ``real" was the claim that quantum theory is a theory of waves {\em in physical space}. This of course is wrong: the state of two particles cannot be expressed as two functions on physical space. Second, a pure-wave formulation misses the essential feature of quantum theory: discreteness, which must be recovered by additional assumptions as there is no reason for a physical wave to have energy related to frequency. Nobody expressed this point clearer than Schr\"odinger himself who much later recognised: ``There was a moment when the creators of wave mechanics [that is, himself] nurtured the illusion of having eliminated the discontinuities in quantum theory.  But the discontinuities eliminated from the equations of the theory reappear the moment the theory is confronted with what we observed."\cite{Schrodinger1996}.   Third, most importantly, if we take $\psi$ to be real  we fall into the infamous ``measurement problem".  In its most vivid form (due to Einstein): if a wave spreads over a region of space, how comes it suddenly concentrates into the single spot where the particle manifests itself?   Schr\"odinger understood the difficulty with his early interpretation, changed his mind repeatedly about the interpretation of the theory \cite{Bitbol1996}, and became one of the most insightful contributors to the debate on the interpretation; but the badly misleading idea of taking the ``quantum state'' as a faithful picture of reality stuck.   

Heisenberg lost the political battle against wave mechanics for a number of reasons: Differential equations are easier to work with than non-commutative algebras. ``Interpretation" wasn't so interesting for many physicists, when the equations of quantum mechanics begun producing wonders.  Dirac found it easier to give the algebra a linear representation, and von Neumann followed: his robust math  brilliantly focused on the algebras, but gave weight to their representation on Hilbert spaces of ``states".  And finally Niels Bohr ---fatherly figure of the  community--- tried to mediate between his two extraordinarily brilliant bickering children, Heisenberg and Schr\"odinger, by obscurely agitating hands about a shamanic ``wave/particle duality".   

A way to get  clarity about quantum mechanics is to undo the conceptual mess raised by Schr\"odinger's introduction of the ``quantum state". This is what the Relational Interpretation does.

\capo{Quantum theory is about physical events, not quantum states}

RQM was proposed in the late Nineties \cite{Rovelli:1995fv}, and has acquired increased clarity over time \cite{Laudisa2001, Smerlak:2006gi, Rovelli2016, Laudisa2017, Rovelli2017b, DiBiagio2020}. Interest in RQM has grown slowly by steadily, attracting attention during the last decade particulary from philosophers \cite{Fraassen:2010fk, pittphilsci3506,Bitbol2010, Dorato2013a, Candiotto2017, Brown2009a,Dorato,Calosi2020a,Oldofredi2021}. 
  
RQM interprets quantum mechanics as a theory about physical \emph{events}, or \emph{facts}.  The theory provides transition amplitudes of the form $W(b,a)$ that determine the probability $P(b,a)=|W(b,a)|^2$ for a fact (or a collection of facts) $b$  to occur, given that a fact  (or a collection of facts) $a$ has occurred.  Facts are the independent variables of the quantum transition amplitudes.  

A fact is for instance a particle having a certain value of a spin component at a certain time, or being at a certain place at a certain time. We perceive and describe the world in terms of such facts.  An example of transition probability is the probability $P(L_\phi=\hbar/2,L_z=\hbar/2)=\cos^2(\phi/2)$ of having spin $\hbar/2$ in a direction at angle $\phi$ from the $z$ axis (a fact) if the spin in the direction $z$ was $\hbar/2$ (a fact).  A fact is quantitatively described by the value of a variable or a set of variables.

Classical mechanics can equally be interpreted as a theory about physical facts, described by values of physical variables (points in phase space). But there are three differences between quantum facts and the corresponding facts of classical mechanics.   First, their dynamical evolution laws are genuinely probabilistic. Second, the spectrum of possible facts is limited by quantum discreteness (for instance: energy or spin can have only certain values). Third, crucially, facts are \emph{sparse} and \emph{relative}. 

Facts are \emph{sparse}: they are realised \emph{only} at the interactions between (any) two physical systems. This is the key physical insight in Heisenberg's seminal paper and a basic assumption of RQM.

Facts are \emph{relative} to the systems that interact.  That is, they are \emph{labelled} by the interacting systems.  This is the core idea of RQM. It gives a general and precise formulation to the central feature of quantum theory, on which Bohr correctly long insisted: contextuality. 
 
The insight of RQM is that the transition amplitudes $W(b,a)$ must be interpreted as determining physical probability amplitudes \emph{only} if the physical facts $a$ and $b$ are \emph{relative to the same systems}. 

If all facts are relative to (or labeled by) the systems involved in the interactions, how come that we can describe a macroscopic world disregarding the labels? The reason decoherence \cite{Zeh1970,Zurek:1981xq,Zurek:1982ii}: because of decoherence, a subset of all relative facts become \emph{stable}  \cite{DiBiagio2020}. This means that if we disregard their labelling we only miss interference effects that are anyway practically unaccessible because of our limited access to the large number of degrees of freedom of the world.  The conventional laboratory ``measurement outcomes" are a particular case of stable facts \cite{DiBiagio2020}; they are relative fact (realised in the pre-measurements) that can be considered stable because of the decoherence due to the interaction of the pointer variable with the environment.

\capo{The relational resolution of the measurement problem}

\paragraph{\bf The problem.}

The measurement problem can be viewed as the apparent contradiction between two postulates of the textbook formulation of quantum theory. On the one hand, the unitary evolution postulate states that the amplitude $W(b(t),a)$ for a fact $b$ to happen at time $t$ changes in $t$ according to the \emph{linear} Schr\"odinger evolution equation $i\hbar\, \partial_t W(b(t),a)=H\, W(b,a)$, where $H$ is a unitary linear operator.  On the other hand, the projection postulate states that probabilities change when a fact occurs.    The contradiction appears because of quantum interference: interference effects in the unitary evolution are cancelled by the projection.   

Explicitly, suppose we know that a fact $a$ has happened and one of $N$ mutually exclusive facts $b_i$ ($i=1\dots N$) can later happen. By composition of probabilities, we expect the probability $P(c)$ for a further fact $c$ to happen to be given by
\begin{equation}
P_{collapse}(c|a)=\sum _i P(c|b_i)P(b_i|a),
\end{equation}
where $P(b|a)$ is the probability for $b$ to happen, given $a$. From the relation between probability and amplitude
\begin{equation}\label{pr}
P_{collapse}(c|a)=\sum _i |W(c,b_i)|^2|W(b_i,a)|^2. 
\end{equation}
But since quantum probabilities are squares of amplitudes and amplitudes sum, linear evolution requires
\begin{eqnarray}
P_{unitary}(c|a)&=& |W(c,a)|^2 = \Big|\sum _i W(c,b_i)W(b_i,a)\Big|^2 \\
&\ne& \sum _i |W(c,b_i)|^2|W(b_i,a)|^2 =P_{collapse}(c|a). \nonumber
\end{eqnarray}
So, what is the probability for $c$ to happen? The projection postulate demands it to be $P_{collapse}(c|a)$ but the unitary evolution postulate requires it to be $P_{unitary}(c|a)$, and the two are different because of interference. 

The textbook answer is that the first holds ``if a measurement has happened", while the second holds if it hasn't.  But what counts as a measurement?  Does Wigner's friend's \cite{Wigner1961} observation count as a measurement? Does Schr\"odinger's cat's  \cite{Schrodinger1935a} observation of the releasing of the poison  count as a measurement?   

A positive answer to these two questions violates the universality of the evolution postulate. But a negative answer (as in the Many World interpretation) fogs the relation between the theory and the facts in terms of which we account for the world. 

\paragraph{\bf The RQM solution.}

The solution offered by RQM is that facts are labelled by systems. They  are labelled by the systems involved in the interaction where the fact happens.  In general, equation \eqref{pr} does not hold if the $b_i$ have different labels than $c$, because the amplitudes $W(b,a)$ determine probabilities \emph{only} if $a$ and $b$ are relative to the same system.  This solves the apparent contradiction. 

For instance, in the Wigner's friend scenario, the friend interacts with a system and a fact is realised \emph{with respect to the friend}. But this fact is not realised with respect to Wigner, who was not involved in the interaction,  and the probability for facts \emph{with respect to Wigner} (realised in interactions with Wigner) still includes interference effects.  A fact with respect to the friend does not count as a fact with respect to Wigner.  With respect to Wigner, it only corresponds to the establishment of an ``entanglement", namely the expectation of a correlation, between the friend and the system.  

In the case of Schr\"odinger's cat, the release or not of the poison is a fact with respect to the cat. An external observer with superior measuring capacities can still detect interference effects (which would not have been possible if the release --or not-- of the poison had become a fact in an interaction with her.) 

Notice that in an ontology based on facts (or events) rather than quantum states, the phrase ``Schr\"o\-dinger's cat is in a quantum superposition" means \emph{only} that we cannot use neither the cat being dead nor the cat being alive as inputs for transition amplitudes.  This is precisely what RQM clarifies: facts are labelled by the systems involved in the interactions and the transition amplitudes $W(b,a)$ have physical meaning only if $a$ and $b$ are relative to a same system. 

\capo{Measurement outcomes and relation with Copenhagen interpretation}

If sufficient decoherence intervenes, the difference between $P_{collapse}(c|a)$ and $P_{unitary}(c|a)$ becomes negligible. In this situation we can safely consider the $b_i$ facts as realised, independently from their labelling.  When we can disregard the labels of a fact, we call it ``stable" \cite{DiBiagio2020}.  This is the case for the textbook laboratory quantum measurement outcomes \cite{DiBiagio2020}, which assume a macroscopic apparatus, and hence decoherence. The macroscopic world is entirely described by stable facts. 

Can we base the ontology \emph{solely} on stable facts? Yes, it is possible. This is what is done in the Copenhagen interpretation, QBism \cite{Fuchs2019} and similar views, such as those in \cite{Grangier2018a,Auffeves2019}.   The prices to be payed with this choice are two. First, decoherence is always approximate only, and only relative to the lack of access to environment degrees of freedom. Therefore the stability of the stable facts is always only approximate. It requires the observer system to have properties that real systems have only approximately. Hence this choice leads to an ontology based on an approximation: no fact is truly exactly stable: any measurement outcome is ultimately like an observation by Wigner's friend: interference with ``the other branch of the state" is alway in principle possible.  We are all actually Wigner's friends, in any measurement.   

The second price to pay is that the resulting ontology does not allow us to conceive reality in the absence of decoherence.  Why should we restrict us to such a narrow view of reality?  Metaphorically: what right does Wigner has to deny the reasonable extrapolation that there are facts relative to his friend, precisely as there are facts relative to himself?  

RQM is based on the observation that enlarging the ontology from stable facts to all relative facts resolves these difficulties, and eliminates the difficulty of characterising what is an observer and what is a measurement.  Any system is an ``observer" in the sense of of being a system with respect to which facts happen.  Decoherence is what characterises ``observers" in the sense of system with resect to which stable facts happen. 

For relative facts, every interaction can be seen as a  ``Copenhagen measurement", but only for the systems involved.  Any physical system can play the role of the ``Copenhagen observer", but only for the facts defined with respect to itself.    From this perspective, RQM is nothing else than a minimal extension of the textbook Copenhagen interpretation, based on the realisation that any physical system can play the role of the  ``observer" and any interaction can play the role of a ``measurement": this is not in contradiction with the permanence of interference through interactions because the ``measured" values are only relative to the interacting systems themselves and do not affect other physical systems. 

In the absence of interactions, there are no events and variables can be genuinely non determinate, as  in the Copenhagen's interpretation \cite{Brown2009a,wood2010everything,Calosi2020a}, because a ``variable" is only a quantity characterising how a system affects another system, and not how a system ``is".

\capo{Meaning of the quantum state}

What is then a ``quantum state"?  In RQM, it is a bookkeeping of known facts, and a tool for predicting the probability of unknown facts, on the basis of the available knowledge.  Since it summarises knowledge about relative facts, the quantum state $\psi$ of a system (and a fortiori its density matrix $\rho$) does not pertain solely to the system.  It pertains also to the other system involved in the interactions that gave rise to the facts considered known.  Hence it is always a relative state.  

The idea that quantum states are relative states, namely states of a physical system relative to a second physical system, is Everett's lasting contribution to the understanding of quantum theory \cite{Everett:1957hd}.   Since $\psi$ is just a theoretical  device we use for bookkeeping information and computing probabilities, it is not a surprise that it jumps if we learn that a fact happens.  The relational interpretation circumvents the PBR (Pusey,  Barrett, Rudolph) theorem because it is not a hidden variable theory \cite{Oldofredi2021}: there is no ``ontological state" at all, because reality is understood in terms of \emph{sparse} relative events. 

A moment of reflection shows that any quantum state used in real laboratories, where scientists use quantum mechanics concretely, is \emph{always} a \emph{relative}  state.  The $\psi$ that physicists use in their laboratories to describe a quantum system is not the hypothetical universal wave function: it is the relative state, in the sense of Everett, that describes the properties of the system, relative to the apparata it is interacting with. 

In a suitable semiclassical approximation we can write $\psi\sim e^{iS}$ where $S$ is a Hamilton-Jacobi function. This shows that the physical nature of $\psi$ is the same as the physical nature of a Hamilton-Jacobi function. Obviously giving $S$  ontological weight in classical mechanics is a mistake: $S$  is not a picture of the ``actual stuff" out there: it is a calculation device used to predict an outcome on the basis of an input.

Quantum mechanics does not need to be interpreted as the theory of the dynamics of a mysterious $\psi$ entity, from which the world of our experience emerges through some involved and obscure argument.  It can simply be interpreted as a theory for computing the probability of facts to occur given that other facts have occurred.  Why should we interpret  $\psi$ differently from $S$, when doing so only creates confusion? 

Finally, there is a simple observation that confirms that it is a mistake to charge the quantum state with ontological weight \cite{Rovelli2016}.  Consider particles with spin $1/2$ that undergo sequences of laboratory measurements of their spin components along different axes. Say that at time $t$ the spin in the $z$ direction of a particle is positive. We can predict that (if nothing else happens in between, and in the absence of any further information) the spin has probability $\cos^2(\phi/2)$ to be up in a direction at angle $\phi$ with the $z$ direction.   {\em This is true irrespectively of which comes earlier between $t$ and $t'$} \cite{Rovelli2016}:   quantum probabilistic predictions are the same forth and back in time \cite{Einstein1931}.   Now: what is the state of the particle during the time interval between $t$ and $t'$?  Answer: it depends on what we consider to be known: if I know the past (respectively, future) value, I can use the state to predict the future (respectively, past) value. This shows manifestly that the state is a coding of our information; not something the particle ``has".  

 It is reasonable to be realist about the values of the spin, which we observe, but not about the $\psi$ in between, {\em because $\psi$ depends on a time orientation, while the observable physics does not}.

\capo{Discreteness}

Several interpretations of quantum theory do not give any importance to discreteness.  Discreteness is not an accessory consequence of quantum theory, it is at its core.  Quantum theory is characterised by the Planck constant $h=2\pi\hbar$. This constant sets the scale of the discreteness of the world and determines how bad is the approximation provided by the continuity of classical mechanics.

Here is a general formulation of quantum discreteness. A fact is quantitively described by the values of a set of variables, that got determined (``measured") in the interaction.  The space of the values of these variables is the \emph{phase space} of the system.  For each degree of freedom, the phase space is two dimensional.  Any measurement has finite precision: it determines a region $R$ of phase space. Classical mechanics assumes that $R$ can be taken to be arbitrarily small. 

The volume $V(R)$ of a phase space region $R$ has dimensions $Length^2\times Mass/Time$ per degree of freedom, namely action. The Planck constant, which has dimensions of an action, fixes is the size of the smallest region that a measurement can determine: 
\begin{align}\label{1.1}
\begin{split}
V(R) &\ge 2\pi\hbar
\end{split}
\end{align}
per degree of freedom.  This is a most general and important physical fact at the core of quantum theory. 

It follows immediately that the number of possible values that a variable distinguishing points within a finite region $R$ of phase space is at most
\begin{align}\label{1.3}
\begin{split}
N\le\frac{V(R)}{2\pi\hbar}
\end{split}
\end{align}
Hence such a variable can take \emph{discrete} values only.  Any variable separating finite regions of phase space is discrete.  

This is the deep root of the most characteristic quantum phenomena: discrete atomic spectra, photons (that is, discrete energy levels of electromagnetic waves), discrete spin, finite black body entropy, finite black hole entropy, etcetera. 

Quantum mechanics gives the values that a physical quantity can take.  Variables are represented by self-adjoint elements $A$ of the non-commutative ($C^*$) algebra defined by $qp-pq=i\hbar$.  The values $a$ that a variable $A(q,p)$ can take are the spectral values of its algebra element $A$, namely the values for which $(a 1-A)$ has no inverse in the algebra.

\capo{Information}

The 1996 seminal RQM  paper \cite{Rovelli:1995fv} indicated \emph{information} as a key concept to understand quantum theory (under the influence of John Wheeler \cite{Wheeler:1990uq,Wheeler:1991fs}) and suggested the program of understanding quantum theory by deriving its peculiar formalism from a transparent set of elementary ``postulates" formulated in terms of information theory.  

The hope was that these would help clarify the meaning of the formalism in the same way in which the two Einstein's Special Relativity postulates clarified the physical content of the Lorentz transformations. 

The two postulates proposed in  \cite{Rovelli:1995fv} are: 

\begin{enumerate}
\item[\bf P1] There is a finite maximal amount of relevant information that can be obtained from a physical system. 

\item[\bf P2] It is always possible to obtain new relevant information from a system by interacting with it.
\end{enumerate}
Here a compact classical phase space is assumed for simplicity. (Any classical system can be approximated by a system with a compact phase space.)   ``Relevant" information means information that contributes to the possibility of predicting the outcome of future interactions. 

The two postulates are not in contradiction with one another because when new information is acquired part of the old information becomes irrelevant.  For instance, measuring the spin of a spin 1/2 particle along a given direction renders the result of any previous spin measurement irrelevant for the probability distribution of future spin measurements.   Related ideas were independently considered by Zeilinger and Brukner \cite{Zeilinger:1999bh,Brukner:2003dq}. 

The first postulate captures the characteristic quantum discreteness. It corresponds to \eqref{1.1}. ``Information" means here nothing else than ``number of possible distinct alternatives".   

The second postulate captures the probabilistic aspects of the theory, because in a deterministic theory there is no way of adding new information once the full information about a system is achieved. 

Historically, the paper  \cite{Rovelli:1995fv} preceded the development of epistemic interpretations like QBism  and the birth of the interpretations of quantum theory based on quantum information \cite{Fuchs1998,Fuchs2001,Spekkens2014}.    It also promoted the program of reconstruction of the formalism of quantum mechanics from physically transparent postulates based on information theory (as a ``theory of principles") \cite{hartle1968quantum,Hardy2001, Grinbaum2005, Spekkens2007,dakic2009quantum,masanes2011derivation,kochen2015reconstruction,oeckl2019local}, which has then grown in a number of different directions.    A particularly successful realisation of the reconstruction program is in the work of H\"ohn \cite{Hoehn2014,Hohn2017}, which uses postulates  based on P1 and P2 above. 

``Information" is understood here in its purely physical sense, namely as correlation: a system has information about another system if the  number of possible states of the two systems is less that the product of the number of possible states of each. For instance  a measuring apparatus that has information about a system after the measure because its pointer variable is correlated to the variable of the system.

The term `information' is ambiguous, with a wide spectrum of meanings ranging from epistemic states of conscious observers all the way to simply counting alternatives, \`a la Shannon. As pointed out by Dorato \cite{Dorato2017}, even in its weakest sense information cannot be taken as a primary notion from which all others can be derived, since it is always information about something.  Nevertheless, information can be a powerful organisational principle in the sense of Einstein's distinction between `principle theories' (like thermodynamics) versus `constructive theories' (like electromagnetism) \cite{Spekkens2014}. The role of the general theory of mechanics is not to list the ingredients of the world ---this is done by the individual mechanical theories, like the Standard Model of particle physics, general relativity, the harmonic oscillator.  The role of the general theory of mechanics (like classical mechanics or quantum mechanics) is to provide a general framework within which specific constructive theories are realized.  From this perspective, the notion of information as number of possible alternatives may play a useful role in accounting for the general structure of the correlations in the physical world. 

It is in this sense that the two postulates can be understood. They are limitations on the structure of the values that variables can take. The list of relevant variables, which define a physical system, and their algebraic relations, are provided by specific quantum theories. 

\capo{The relation between formalism and interpretation}

The non-commutativity of the physical variables is Heisenberg technical breakthrough, understood and formalised by Born and Jordan, and independently by Dirac. Both Dirac and the G\"ottingen group arrived independently at the main equation \eqref{1.2}
(and its generalisations).  

Heisenberg's breakthrough is the idea of keeping the same equations as in the classical theory, but replacing commuting variables with non commuting ones, satisfying (as later realized by Born and Jordan) equation \eqref{1.2}.  

In a sense, one could say that quantum theory has the same equations as the classical theory, plus this single equation.  The formalism of quantum theory is condensed in this single equation.  The entire quantum phenomenology follows from this one equation. 

As well known, the Heisenberg uncertainty relations
\be
\Delta q\ \Delta p\le \hbar/2
\ee
can be derived from \eqref{1.2} in a few lines. Hence \eqref{1.2} leads immediately to \eqref{1.3}, hence to discreteness, establishing the quantitative aspect of discreteness (by $\hbar$).  This is directly related to the first postulate.    

On the other hand, the non-commutativity  expressed in \eqref{1.2} reflects the fact that the result of the measurements of $q$ and $p$ depends on the order in which measurements are made: this is  what blocks the possibility of a complete specification of all the variables of a system: it is directly related to the second postulate.  Both postulates are made concrete in the math by equation  \eqref{1.2}. 

In other words, the non-commutativity of the algebra of the variables (measured by $\hbar$) is the mathematical expression of the physical fact that variables cannot be  simultaneously sharp, hence there is a ($\hbar$-size) minimal volume attainable in phase space, and predictions are probabilistic. 

The fact that values of variables can be predicted only probabilistically raises immediately the key interpretational question of quantum mechanics: when and how is a probabilistic disposition resolved into an actual value? RQM  has a simple answer: when any two systems interact, provided that we label the resulting facts with the interacting systems themselves.

\capo{Other considerations} 

\begin{itemize}
\item{\em No-go theorems for non-relative facts.}
There are a number of no-go theorems for non-relative (absolute) facts  \cite{Frauchiger2018a,Brukner2018,Bong2020}, and some experimental confirmations of them \cite{Proietti2019}. 
The existence of facts not labelled by systems is among the inputs of these theorems, hence these no-go theorems can be taken as direct evidence in favour of RQM. See a detailed discussion in \cite{DiBiagio2020} and also \cite{Waaijer2019,Cavalcanti2019}. 

\item{\em Locality.}
The way in which quantum non locality is realised in RQM \cite{Smerlak:2006gi, Laudisa2017a} has been clarified recently in \cite{Martin-Dussaud2019a,Pienaar2019}, showing how the tension with relativity is alleviated by the fact that a measurement in a location cannot be an element of reality with respect to a distantly located observer. 

\item{\em Quantum Gravity.}
RQM is fit for quantum gravity \cite{Rovelli:2004fk, Rovelli:2014ssa} (in fact, this was its historical motivation). In quantum gravity, we do not have a background spacetime where to locate things.  Quantum gravity works because the quantum relationalism emphasized by RQM combines in a surprisingly natural manner with the relationalism of general relativity.  Locality is what makes this work \cite{Vidotto2013}: the quantum mechanical notion of ``physical system" an be identified with the general relativistic notion of ``spacetime region".  The quantum mechanical notion of ``interaction" between systems is identified with the general relativistic notion of ``adjacency" between spacetime regions.  Locality assures that interaction requires (and defines) adjacency.  
Thus quantum ``events" can be associated to three dimensional surfaces bounding spacetime regions and  quantum mechanical transition amplitudes are associated to ``processes" identified with the spacetime regions themselves. In other words, variables actualise at three dimensional boundaries, with respect to (arbitrary) spacetime partitions.   The theory can then be used locally, without necessarily assuming anything about the global aspects of the universe.

Some researchers think that the only observables available in quantum gravity are asymptotic observables at infinite distance from the system, because decoherence requires an infinite number of degrees of freedom  (see for instance \cite{Arkani}).  RQM offers a way to talk about transition probabilities also in the absence of (perfect) decoherence, circumventing the problem, therefore allowing meaningful observables for processes in finite spacetime regions. 

\end{itemize}

\capo{Philosophical implications}

The beauty of the problem of the interpretation of quantum mechanics is the fact that the spectacular and unmatched empirical success of the theory forces us to give up at least \emph{some} cherished philosophical assumption.   Which one is convenient to give up is the open question. 

The relational interpretation offers an alternative to the quantum state realism of Many-Words-like interpretation and to the strong instrumentalism of the strictly epistemic interpretations  \cite{Zeilinger:1999bh, Hardy2001, Brukner:2003dq, Spekkens2007, Hohn2017a, Fuchs2019}.  It avoids introducing ``many worlds", hidden variables, physical collapse, and also avoids the instrumentalism of other epistemic interpretations.  But, like any other consistent interpretation of quantum theory, it comes at a price. 

It is compatible with  diverse philosophical perspectives (see below). But not all.  Its main cost is a challenge to a strong version of realism, which is implied by its radical relational stance.  

Relationality is no surprise in physics.  In classical mechanics the velocity of an object has no meaning by itself: it is only defined with respect to another object.   The color of a quark in  strong-interaction theory has no meaning by itself: only the relative color of two quarks has meaning.   In electromagnetism, the potential at a point has no meaning, unless another point is taken as reference; that is, only relative potentials have meanings. In general relativity, the location of something is only defined with respect to the gravitational field, or with respect to other physical entities; and so on.   But quantum theory takes this ubiquitous relationalism, to a new level: the actual value of \emph{all} physical quantities of \emph{any} system is only meaningful in relation to another system.   Value actualisation is a relational notion like velocity.  

Hence the conceptual cost of RQM is giving up a strong form of realism: not only to give up the assumption that physical variables take values at all times, but also to accept that they take values at different times for different systems.

Strong realism is ingrained in our common-sense view of the world, and is often given for granted.   For instance it is among the hidden hypotheses of the Pusey-Barrett-Rudolph theorem \cite{Pusey}. The relational interpretation circumvents theorems like these because these \emph{assume} that at every moment of time all properties are well defined. (For a review, see \cite{Leifer2014}.)  This assumption is explicitly denied in relational QM: properties do not exist at all times: they are properties of events and the events happen at interactions. 
On the same vein, in \cite{Laudisa2017a} Laudisa criticises relational QM because it does not provide a ``deeper justification" for the ``state reduction process".  It is a stance based on a very strong realist (in the narrow sense specified above) philosophical assumption.   

A second element in RQM that challenges strong realism is that values taken with respect to different systems can be compared \cite{Rovelli:1995fv} (hence there no solipsism), but the comparison amounts to a physical interaction, and its sharpness is limited by $\hbar$. Therefore we cannot escape from the limitation to partial views: there is no coherent global view available.   Matthew Brown has discussed this point in \cite{Brown2009a}. 

The third element in RQM that challenges strong realism, emphasized by Dorato \cite{Dorato}, is the `anti-monistic' stance implicit in relational QM. Since the state of a system is a bookkeeping device of interactions with \emph{something else}, it follows immediately that there is no meaning in ``the quantum state of the full universe". There is no something else to the universe. Everett's relative states are the only quantum states we can meaningfully talk about. Every quantum state is an Everett's quantum state.   (This does not prevent conventional quantum cosmology to be studied, since physical cosmology is not the science of everything: it is the science of the largest-scale degrees of freedom \cite{Vidotto2017}. They are well defined and stable, relative to our observations.) 

In the philosophical literature RQM as been extensively discussed by Bas van Fraassen  \cite{Fraassen:2010fk} 
from a marked empiricist perspective, by Michel Bitbol \cite{pittphilsci3506,Bitbol2010} who has given a neo-Kantian version of the interpretation, by Mauro Dorato \cite{Dorato2013a} who has defended it against a number of potential objections and discussed its philosophical implication on monism and dispositionalism, and recently by Laura Candiotto \cite{Candiotto2017} who has given it an intriguing reading in terms of (Ontic) Structural Realism \cite{French2011}.  Metaphysical and epistemological implications of relational QM have also been discussed by  Matthew Brown \cite{Brown2009a} and Daniel Wolf (n\'e Wood) \cite{wood2010everything}. 

RQM has aspects in common with QBism \cite{Fuchs2014}, with Healey’s pragmatist approach \cite{Healey1989}  and is close in spirit with the view of quantum theory discussed by Zeilinger and Bruckner  \cite{Zeilinger:1999bh,Brukner:2003dq}.  There are similarities 	with recent ideas by
Auff\`eves and Grangier \cite{Grangier2018a,Auffeves2019}.

\capo{Perspective}

Relational QM is a radical attempt to cash out the breakthrough that originated the theory: the world is described by facts described by values of variables that obey the equations of classical mechanics, \emph{but} products of these variable have a tiny non-commutativity that generically prevents sharp value assignment, leading to \emph{discreteness}, \emph{probability} and to the {contextual}, \emph{relational} character of value assignment.  

The founders expressed this contextual character on Nature in the ``observer-measurement" language. This language requires that special systems (the observer, the classical world, macroscopic objects...) escape the quantum limitations.  But nothing of that sort (and in particular no ``subjective states of conscious observers") is needed in the interpretation of QM.   We can relinquish this exception, and realise that \emph{any} physical system can play the role of a Copenhagen's ``observer". Relational QM is Copenhagen quantum mechanics made democratic by bringing all systems onto the same footing.   Macroscopic observers, that loose information to decoherence, can forget the labelling of facts.  

In the history of physics progress has often happened by realising that some naively realist expectations were ill founded, and therefore by dropping corresponding questions: How are the spheres governing the orbits of planet arranged? What is the mechanical underpinning of the electric and magnetic fields? Into where is the universe expanding? To some extent, one can say that modern science itself was born in Newton's celebrated ``hypotheses non fingo", which is precisely the recognition that questions of this sort might be misleading.  

When everybody else was trying to find dynamical laws accounting for atoms, Heisenberg's breakthrough was to realise that the known laws where already good enough, but the question of the actual continuous orbit of the electron was ill posed: the world is better comprehensible in terms of a sparse relational ontology. RQM is the realisation that this is what we have learned about the world with quantum physics. 

\vspace{1cm}
\centerline{***}
{A special thank to Andrea Di Biagio and Guido Baciagaluppi for corrections, suggestions and guidance.}


\begin{thebibliography}{10}

\bibitem{Schrodinger:1926fk}
E.~Schr{\"{o}}dinger, ``{Quantisierung als Eigenwertproblem (Erste
  Mitteilung)},'' {\em Annalen der Physik} {\bf 79} (1926)  361--76.

\bibitem{Pauli1926}
W.~Pauli, ``{{\"{U}}ber das Wasserstoffspektrum vom Standpunkt der neuen
  Quantenmechanik [On the hydrogen spectrum from the standpoint of the new
  quantum mechanics]},'' {\em Zeitschrift f{\"{u}}r Physik} {\bf 36} (1926)
  336--363.

\bibitem{Heisenberg1925a}
W.~Heisenberg, ``{Uber quantentheoretische Umdeutung kinematischer und
  mechanischer Beziehungen.},'' {\em Zeitschrift f{\"{u}}r Physik} {\bf 33}
  (1925) no.~1, 879--893.

\bibitem{Born1925b}
M.~Born and P.~Jordan, ``{Zur Quantenmechanik},'' {\em Z. Phys.} {\bf 34}
  (1925)  854--888.

\bibitem{Born1926}
M.~Born, P.~Jordan, and W.~Heisenberg, ``{Zur Quantenmechanik II},'' {\em
  Zeitschrift f{\"{u}}r Physik} {\bf 35} (1926)  557--615.

\bibitem{Dirac1925}
P.~A.~M. Dirac, ``{The fundamental equations of quantum mechanics},'' {\em c.
  London, Ser. A} {\bf 645-653} (1925)  .

\bibitem{Fedak2009}
W.~a. Fedak and J.~J. Prentis, ``{The 1925 Born and Jordan paper 'On quantum
  mechanics'},'' \href{http://dx.doi.org/10.1119/1.3009634}{{\em American
  Journal of Physics} {\bf 77} (2009)  128}.

\bibitem{Waerden1967}
B.~van~der Waerden, {\em {Sources of quantum mechanics}}.
\newblock North Holland, 1967.

\bibitem{Kumar2008}
M.~Kumar, {\em {Quantum: Einstein, Bohr and the Great Debate About the Nature
  of Reality}}.
\newblock Icon Books Ltd, 2008.

\bibitem{Schrodinger1996}
E.~Schr{\"{o}}dinger, {\em {Nature and the Greeks and Science and Humanism}}.
\newblock Cambridge Univeristity Press, Cambridge, 1996.

\bibitem{Bitbol1996}
M.~Bitbol, ``{Schr{\"{o}}dinger's Philosophy of Quantum Mechanics (Boston
  Studies in the Philosophy and History of Science)},''
\newblock Springer, Berlin, 1996.

\bibitem{Rovelli:1995fv}
C.~Rovelli, ``{Relational Quantum Mechanics},'' {\em Int. J. Theor. Phys.} {\bf
  35} (1996)  1637, \href{http://arxiv.org/abs/9609002}{{\tt arXiv:9609002
  [quant-ph]}}.

\bibitem{Laudisa2001}
F.~Laudisa, ``{The EPR argument in a relational interpretation of quantum
  mechanics},'' {\em Foundations of Physics Letters} {\bf 14} (2001)  119--132,
  \href{http://arxiv.org/abs/0011016v1}{{\tt arXiv:0011016v1 [quant-ph]}}.

\bibitem{Smerlak:2006gi}
M.~Smerlak and C.~Rovelli, ``{Relational EPR},'' {\em Found. Phys.} {\bf 37}
  (2007)  427--445, \href{http://arxiv.org/abs/0604064}{{\tt arXiv:0604064
  [quant-ph]}}.

\bibitem{Rovelli2016}
C.~Rovelli, ``{An Argument Against the Realistic Interpretation of the Wave
  Function},'' {\em Foundations of Physics} {\bf 46} (2016) no.~10, 1229--1237,
  \href{http://arxiv.org/abs/1508.05533}{{\tt arXiv:1508.05533}}.

\bibitem{Laudisa2017}
F.~Laudisa and C.~Rovelli, ``{Stanford Encyclopedia of Philosophy},'' 2017.
\newblock \url{https://plato.stanford.edu/entries/qm-relational/}.

\bibitem{Rovelli2017b}
C.~Rovelli, ``{`Space is blue and birds fly through it'},'' {\em Philosophical
  Transactions A} (2017)  , \href{http://arxiv.org/abs/1712.02894}{{\tt
  arXiv:1712.02894}}.

\bibitem{DiBiagio2020}
A.~{Di Biagio} and C.~Rovelli, ``{Stable Facts, Relative Facts},''
  \href{http://arxiv.org/abs/2006.15543}{{\tt arXiv:2006.15543}}.

\bibitem{Fraassen:2010fk}
B.~C. van Fraassen, ``{Rovelli's World},'' {\em Foundations of Physics} {\bf
  40} (2010) no.~4, 390--417.
  \url{https://www.princeton.edu/~fraassen/abstract/Rovelli_sWorld-FIN.pdf}.

\bibitem{pittphilsci3506}
M.~Bitbol, ``{Physical Relations or Functional Relations ? A non-metaphysical
  construal of Rovelli's Relational Quantum Mechanics},'' sep, 2007.
\newblock \url{http://philsci-archive.pitt.edu/3506/}.

\bibitem{Bitbol2010}
M.~Bitbol, {\em {De l'int{\'{e}}rieur du monde. (Relational Quantum Mechanics
  is extensively discussed in Chapter 2)}}.
\newblock Flammarion, Paris, 2010.

\bibitem{Dorato2013a}
M.~Dorato, ``{Rovelli' s relational quantum mechanics, monism and quantum
  becoming},'' in {\em The Metaphysics of Relations}, A.~Marmodoro and
  A.~Yates, eds., pp.~290--324.
\newblock Oxford University Press, 2016.
\newblock \href{http://arxiv.org/abs/1309.0132}{{\tt arXiv:1309.0132}}.

\bibitem{Candiotto2017}
L.~Candiotto, ``{The reality of relations},'' {\em Giornale di Metafisica} {\bf
  2} (2017)  537--551.

\bibitem{Brown2009a}
M.~J. Brown, ``{Relational Quantum Mechanics and the Determinacy Problem},''
  {\em British Journal for the Philosophy of Science} {\bf LX} (2009) no.~4,
  679--695.

\bibitem{Dorato}
M.~Dorato, ``{Bohr meets Rovelli:a dispositionalist account of the quantum
  state},'' jun, 2020.
 
  \bibitem{Calosi2020a}
C.~Calosi, M.~Mariani , ``{Quantum relational indeterminacy},''
  {\em Studies in History and Philosophy of Science Part B} {\bf
  71} (2020)  158--169.

  \bibitem{Oldofredi2021}
A.~Oldofredi, C.~Calosi, ``{Relational Quantum Mechanics and the PBR Theorem: A Peaceful Coexistence},''
  {\em Foundations of Physics} {\bf
4} (2021)  158--169.
\newblock \href{http://arxiv.org/abs/2107.02566}{{\tt arXiv:2107.02566}}.

\bibitem{Zeh1970}
H.~D. Zeh, ``{On the Interpretation of Measurement in QuantumTheory},'' {\em
  Foundations of Physics} {\bf 1} (1970)  79.

\bibitem{Zurek:1981xq}
W.~H. Zurek, ``{Pointer Basis of Quantum Apparatus: Into What Mixture Does the
  Wave Packet Collapse?},'' {\em Phys. Rev.} {\bf D24} (1981)  1516--1525.

\bibitem{Zurek:1982ii}
W.~H. Zurek, ``{Environment induced superselection rules},'' {\em Phys. Rev.}
  {\bf D26} (1982)  1862--1880.

\bibitem{Wigner1961}
P.~E. Wigner, ``{Remarks on the mind-body question},'' in {\em The Scientists
  Speculates}, I.~Good, ed., pp.~248--302.
\newblock Heinemann, London, 1961.
\newblock
  \url{http://www.informationphilosopher.com/solutions/scientists/wigner/Wigner_Remarks.pdf}.

\bibitem{Schrodinger1935a}
E.~Schr{\"{o}}dinger, ``{Die gegenw{\"{a}}rtige Situation in der
  Quantenmechanik},''{\em Die Naturwissenschaften} {\bf 23} (nov, 1935)
  807--812.

\bibitem{Fuchs2019}
C.~A. Fuchs and B.~C. Stacey, ``{QBism: Quantum theory as a hero's handbook},''
  in {\em Proceedings of the International School of Physics "Enrico Fermi"},
  vol.~197, pp.~133--202.
\newblock dec, 2019.
\newblock \href{http://arxiv.org/abs/1612.07308}{{\tt arXiv:1612.07308}}.

\bibitem{Grangier2018a}
A.~Auff{\`{e}}ves and P.~Grangier, ``{What is quantum in quantum
  randomness?},''{\em Philosophical Transactions of the Royal Society A:
  Mathematical, Physical and Engineering Sciences} {\bf 376} (apr, 2018)  ,
  \href{http://arxiv.org/abs/1804.04807}{{\tt arXiv:1804.04807}}.

\bibitem{Auffeves2019}
A.~Auff{\`{e}}ves and P.~Grangier, ``{A Generic Model for Quantum
  Measurements},''{\em Entropy} {\bf 21} (jul, 2019)  904,
  \href{http://arxiv.org/abs/1907.11261}{{\tt arXiv:1907.11261}}.



\bibitem{wood2010everything}
D.~Wood, ``{Everything Is Relative: {\{}{\{}Has Rovelli{\}}{\}} Found the Way
  out of the Woods?},'' may, 2010.

\bibitem{Everett:1957hd}
H.~Everett, ``{Relative state formulation of quantum mechanics},'' {\em Rev.
  Mod. Phys.} {\bf 29} (1957)  454--462.

\bibitem{Einstein1931}
A.~Einstein, R.~Tolman, and B.~Podolsky, ``{Knowledge of past and future in
  quantum mechanics},'' {\em Phys. Rev.} {\bf 37} (1931)  780.

\bibitem{Wheeler:1990uq}
J.~A. Wheeler, ``{Information, physics, quantum: The search for links},'' in
  {\em Complexity, Entropy, and the Physics of Information}, W.~H. Zurek, ed.,
  pp.~3--28.
\newblock Addison-Wesley, Redwood City, California, 1990.

\bibitem{Wheeler:1991fs}
J.~A. Wheeler, ``{It from bit},'' in {\em Moscow 1991, Proceedings, Sakharov
  memorial lectures in physics}, vol.~2, p.~751.
\newblock 1991.

\bibitem{Zeilinger:1999bh}
A.~Zeilinger, ``{A Foundational Principle for Quantum Mechanics},'' {\em Found.
  Phys} {\bf 29} (1999)  631--643.

\bibitem{Brukner:2003dq}
C.~Brukner and A.~Zeilinger, ``{Information and fundamental elements of the
  structure of quantum theory},'' in {\em Time, Quantum, Information},
  L.~Castell and O.~Ischebeck, eds., pp.~323--354.
\newblock Springer: Berlin Heidelberg, 2003.
\newblock \href{http://arxiv.org/abs/0212084}{{\tt arXiv:0212084 [quant-ph]}}.

\bibitem{Fuchs1998}
C.~A. Fuchs, ``{Information Gain vs. State Disturbance in Quantum Theory},''
  {\em Fortschritte der Physik} {\bf 46} (1998)  535----565.

\bibitem{Fuchs2001}
C.~A. Fuchs, ``{Quantum Foundations in the Light of Quantum Information},'' in
  {\em Decoherence and Its Implications in Quantum Computation and Information
  Transfer: Proceedings of the NATO Advanced Research Workshop}, A.~Gonis and
  P.~E.~A. Turchi, eds.
\newblock Amsterdam, ios press~ed., 2001.
\newblock \href{http://arxiv.org/abs/0106166}{{\tt arXiv:0106166 [quant-ph]}}.

\bibitem{Spekkens2014}
R.~Spekkens, ``{The invasion of Physics by Information Theory, talk at
  Perimeter Institute, March 26, 2014.},''. \url{http://pirsa.org/14030085/}.

\bibitem{hartle1968quantum}
J.~B. Hartle, ``{Quantum Mechanics of Individual Systems},''
  \href{http://arxiv.org/abs/1907.02953}{{\tt arXiv:1907.02953}}.
  \url{https://aapt.scitation.org/doi/10.1119/1.1975096}.

\bibitem{Hardy2001}
L.~Hardy, ``{Quantum Theory From Five Reasonable Axioms},''
  \href{http://arxiv.org/abs/0101012}{{\tt arXiv:0101012 [quant-ph]}}.

\bibitem{Grinbaum2005}
A.~Grinbaum, ``{Information-theoretic princple entails orthomodularity of a
  lattice},'' {\em Foundations of Physics Letters} {\bf 18} (2005)  563--572,
  \href{http://arxiv.org/abs/0509106}{{\tt arXiv:0509106 [quant-ph]}}.

\bibitem{Spekkens2007}
R.~W. Spekkens, ``{Evidence for the epistemic view of quantum states: A toy
  theory},''{\em Physical Review A - Atomic, Molecular, and Optical Physics}
  {\bf 75} (mar, 2007)  032110.

\bibitem{dakic2009quantum}
B.~Dakic and C.~Brukner, ``{Quantum Theory and Beyond: Is Entanglement
  Special?},'' \href{http://arxiv.org/abs/0911.0695}{{\tt arXiv:0911.0695
  [quant-ph]}}. \url{http://arxiv.org/abs/0911.0695}.

\bibitem{masanes2011derivation}
L.~Masanes and M.~P. M{\"{u}}ller, ``{A Derivation of Quantum Theory from
  Physical Requirements},''.
  \url{https://doi.org/10.1088{\%}2F1367-2630{\%}2F13{\%}2F6{\%}2F063001}.

\bibitem{kochen2015reconstruction}
S.~Kochen, ``{A Reconstruction of Quantum Mechanics},''
  \href{http://arxiv.org/abs/1306.3951}{{\tt arXiv:1306.3951 [quant-ph]}}.
  \url{http://arxiv.org/abs/1306.3951}.

\bibitem{oeckl2019local}
R.~Oeckl, ``{A Local and Operational Framework for the Foundations of
  Physics},'' \href{http://arxiv.org/abs/1610.09052}{{\tt arXiv:1610.09052}}.
  \url{http://arxiv.org/abs/1610.09052}.

\bibitem{Hoehn2014}
P.~A. H{\"{o}}hn, ``{Toolbox for reconstructing quantum theory from rules on
  information acquisition},'' {\em Quantum} {\bf 1} (2017) no.~38, ,
  \href{http://arxiv.org/abs/1412.8323}{{\tt arXiv:1412.8323}}.

\bibitem{Hohn2017}
P.~A. H{\"{o}}hn and C.~S. Wever, ``{Quantum theory from questions},'' {\em
  Physical Review A} {\bf 95} (2017)  ,
  \href{http://arxiv.org/abs/1511.01130}{{\tt arXiv:1511.01130}}.

\bibitem{Dorato2017}
M.~Dorato, ``{Dynamical versus structural explanations in scientific
  revolutions},'' {\em Synthese} {\bf 194} (2017)  2307--2327.
  \url{http://philsci-archive.pitt.edu/10982/}.

\bibitem{Frauchiger2018a}
D.~Frauchiger and R.~Renner, ``{Quantum theory cannot consistently describe the
  use of itself},''{\em Nature Communications} {\bf 9} (apr, 2018)  ,
  \href{http://arxiv.org/abs/1604.07422}{{\tt arXiv:1604.07422}}.

\bibitem{Brukner2018}
{\v{C}}.~Brukner, ``{A No-Go theorem for observer-independent facts},''{\em
  Entropy} {\bf 20} (apr, 2018)  .

\bibitem{Bong2020}
K.~W. Bong, A.~Utreras-Alarc{\'{o}}n, F.~Ghafari, Y.~C. Liang, N.~Tischler,
  E.~G. Cavalcanti, G.~J. Pryde, and H.~M. Wiseman, ``{A strong no-go theorem
  on the Wigner's friend paradox},''{\em Nature Physics} (jul, 2020)  ,
  \href{http://arxiv.org/abs/1907.05607}{{\tt arXiv:1907.05607}}.

\bibitem{Proietti2019}
M.~Proietti, A.~Pickston, F.~Graffitti, P.~Barrow, D.~Kundys, C.~Branciard,
  M.~Ringbauer, and A.~Fedrizzi, ``{Experimental test of local observer
  independence},'' {\em Science Advances} {\bf 5} (2019) no.~9, ,
  \href{http://arxiv.org/abs/1902.05080}{{\tt arXiv:1902.05080}}.

\bibitem{Waaijer2019}
M.~Waaijer and J.~van Neerven, ``{Relational analysis of the Frauchiger--Renner
  paradox and interaction-free detection of records from the past},''
  \href{http://arxiv.org/abs/1902.07139}{{\tt arXiv:1902.07139}}.

\bibitem{Cavalcanti2019}
E.~Cavalcanti and H.~M.~Wiseman ``{Implications of Local Friendliness violation to quantum
  causality},'' {\em  Entropy}, {\bf 23}, 925.
   \href{http://arxiv.org/abs/2106.04065}{{\tt arXiv:2106.04065}}.

\bibitem{Laudisa2017a}
F.~Laudisa, ``{Open Problems in Relational Quantum Mechanics},''
  \href{http://arxiv.org/abs/1710.07556}{{\tt arXiv:1710.07556}}.

\bibitem{Martin-Dussaud2019a}
P.~Martin-Dussaud, C.~Rovelli, and F.~Zalamea, ``{The Notion of Locality in
  Relational Quantum Mechanics},'' {\em Foundations of Physics} {\bf 49} (2019)
  no.~2, 96--106, \href{http://arxiv.org/abs/1806.08150}{{\tt
  arXiv:1806.08150}}.

\bibitem{Pienaar2019}
J.~Pienaar, ``{Comment on ‚??The Notion of Locality in Relational Quantum
  Mechanics"},'' {\em Foundations of Physics} {\bf 49} (2019)  1404--1414,
  \href{http://arxiv.org/abs/1807.06457}{{\tt arXiv:1807.06457}}.

\bibitem{Rovelli:2004fk}
C.~Rovelli, {\em {Quantum Gravity}}.
\newblock Cambridge University Press, 2004.

\bibitem{Rovelli:2014ssa}
C.~Rovelli and F.~Vidotto, {\em {Covariant Loop Quantum Gravity}}.
\newblock Cambridge University Press, 2014.

\bibitem{Vidotto2013}
F.~Vidotto, ``{Atomism and Relationism as guiding principles for Quantum
  Gravity},'' in {\em Seminar on the Philosophical Foundation of Quantum
  Gravity, Chicago 26-28 Sept. 2013}, vol.~FFP14, p.~222.
\newblock SISSA, 2013.
\newblock \url{https://inspirehep.net/record/1487207}.

\bibitem{Arkani}
Nina Arkani-Hamed, interview in \emph{Conversations on Quantum Gravity}, pp. 33–62. Cambridge Univeristity Press 2021.

\bibitem{Hohn2017a}
P.~A. H{\"{o}}hn and C.~S. Wever, ``{Quantum theory from questions},''{\em
  Physical Review A} {\bf 95} (jan, 2017)  012102,
  \href{http://arxiv.org/abs/1511.01130}{{\tt arXiv:1511.01130}}.

\bibitem{Pusey}
M.~Pusey, J.~Barrett, and T.~Rudolph, ``{On the reality of the quantum
  state},'' {\em Nature Physics} {\bf 8} (2012)  475--478,
  \href{http://arxiv.org/abs/1111.3328}{{\tt arXiv:1111.3328}}.

\bibitem{Leifer2014}
M.~S. Leifer, ``{Is the quantum state real? An extended review of
  $\psi$-ontology theorems},'' \href{http://arxiv.org/abs/1409.1570}{{\tt
  arXiv:1409.1570}}.

\bibitem{Vidotto2017}
F.~Vidotto, ``{Relational quantum cosmology},'' in {\em The Philosophy of
  Cosmology}, pp.~297--316.
\newblock aug, 2017.
\newblock \href{http://arxiv.org/abs/1508.05543}{{\tt arXiv:1508.05543}}.

\bibitem{French2011}
S.~French and J.~Ladyman, ``{In Defence of Ontic Structural Realism},'' in {\em
  Scientific Structuralism}, A.~Bokulich and P.~Bokulich, eds., pp.~25--42.
\newblock Springer, Dordrecth, 2011.

\bibitem{Fuchs2014}
C.~Fuchs, N.~Mermin, and R.~Schack, ``{An introduction to QBism with an
  application to the locality of quantum mechanics},'' {\em Am. J. Phys.} {\bf
  82} (2014)  749--754.

\bibitem{Healey1989}
R.~Healey, {\em {The Quantum Revolution in Philosophy}}.
\newblock Oxford University Press, Oxford, 2016.

  \end{thebibliography}

\end{document}